\documentclass{appolb}
\usepackage{graphicx}

\begin{document}
\title{Strong and Electromagnetic Forces in Heavy Ion Collisions%
\thanks{Presented at the International Symposium on Multiparticle Dynamics - ISMD2012, Kielce, 16-21 September 2012.}%
}
\author{Mariola K\l{}usek-Gawenda$^1$, 
Ewa Kozik$^1$,
Andrzej Rybicki$^1$, 
Iwona Sputowska$^1$, 
Antoni Szczurek$^{1,2}$
\address{$^1$H.~Niewodnicza\'{n}ski Institute of Nuclear Physics, Polish 
 Academy of Sciences, Radzikowskiego 152, 31-342~Krak\'ow, Poland\\
 $^2$University of Rzesz\'ow, Rejtana 16, 35-959 Rzesz\'ow, Poland}
}
\maketitle
\begin{abstract}
 The interplay between the strong and electromagnetic force in high 
energy nucleus-nucleus collisions was studied experimentally and 
theoretically in our earlier works. This effect appeared 
to result in very large distortions in spectra of charged pions produced 
in the collision. It was also found to bring new, independent 
information on the space-time evolution of the non-perturbative process 
of particle production. 
 In this paper, we present our new results on 
the influence of the spectator-induced electromagnetic force on spectra 
of charged particles produced in two different Pb-induced reactions. For 
the 
first time, we 
also address 
the topic of p+A 
collisions in view of obtaining information about their centrality and 
nuclear break-up, both subjects being of importance
 in the context of the
new p+A data collected at the LHC.
\end{abstract}
  
\section{Introduction}
 This paper addresses a specific class of nucleus-nucleus 
collisions, where a large ``spectator system'' survives.
 Such specific reactions give the opportunity to investigate the 
interplay between phenomena occurring in the participant and spectator 
zones. In particular, this is the case for our study which concerns the 
interplay between the strong and electromagnetic forces in such 
reactions. 
This 
 can be used as a 
new source of information on the collision dynamics.

\begin{figure}[htb]
\centerline{%
\includegraphics[width=5.7cm]{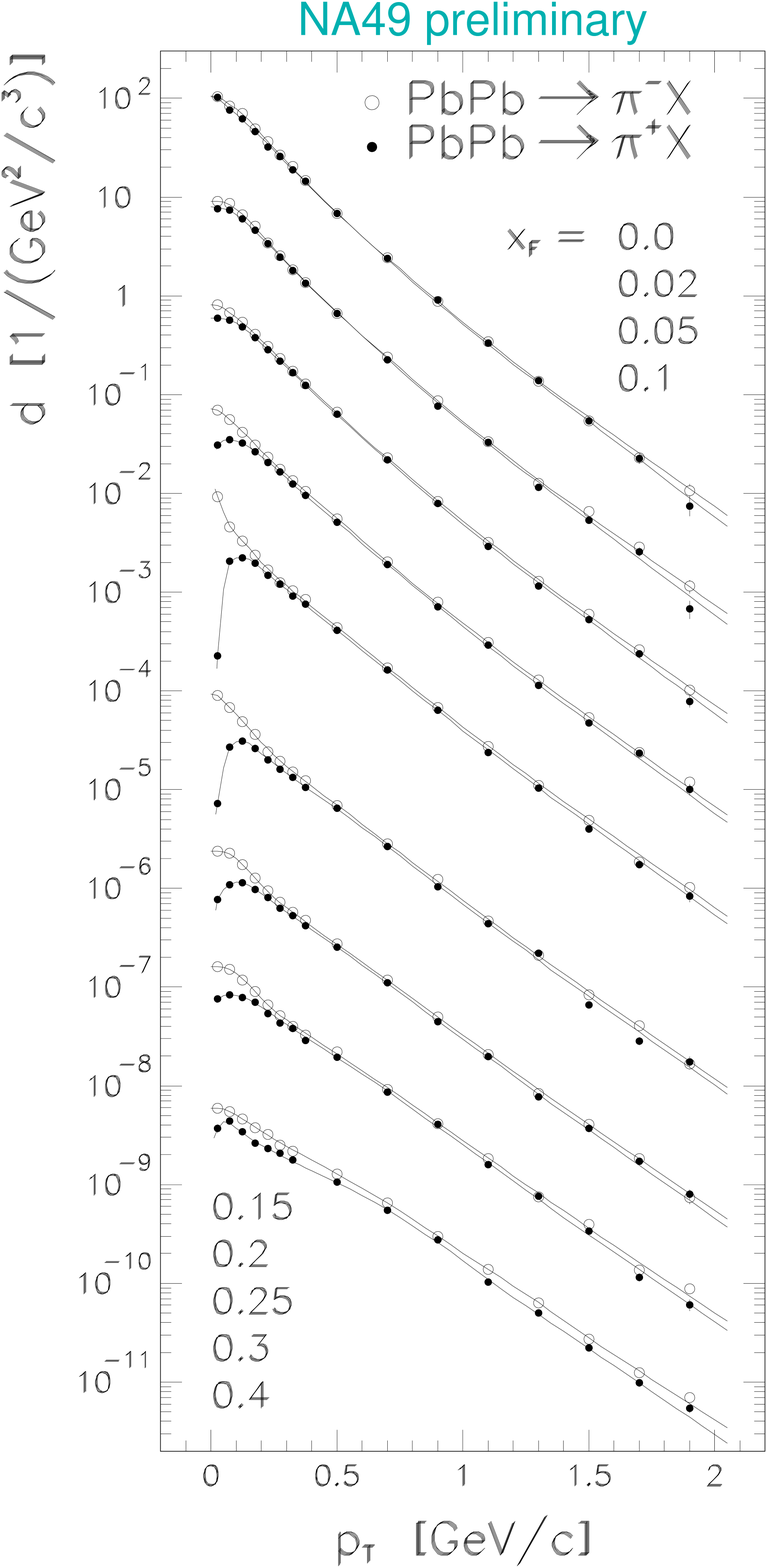}
\hspace*{0.4cm}
\includegraphics[width=5.7cm]{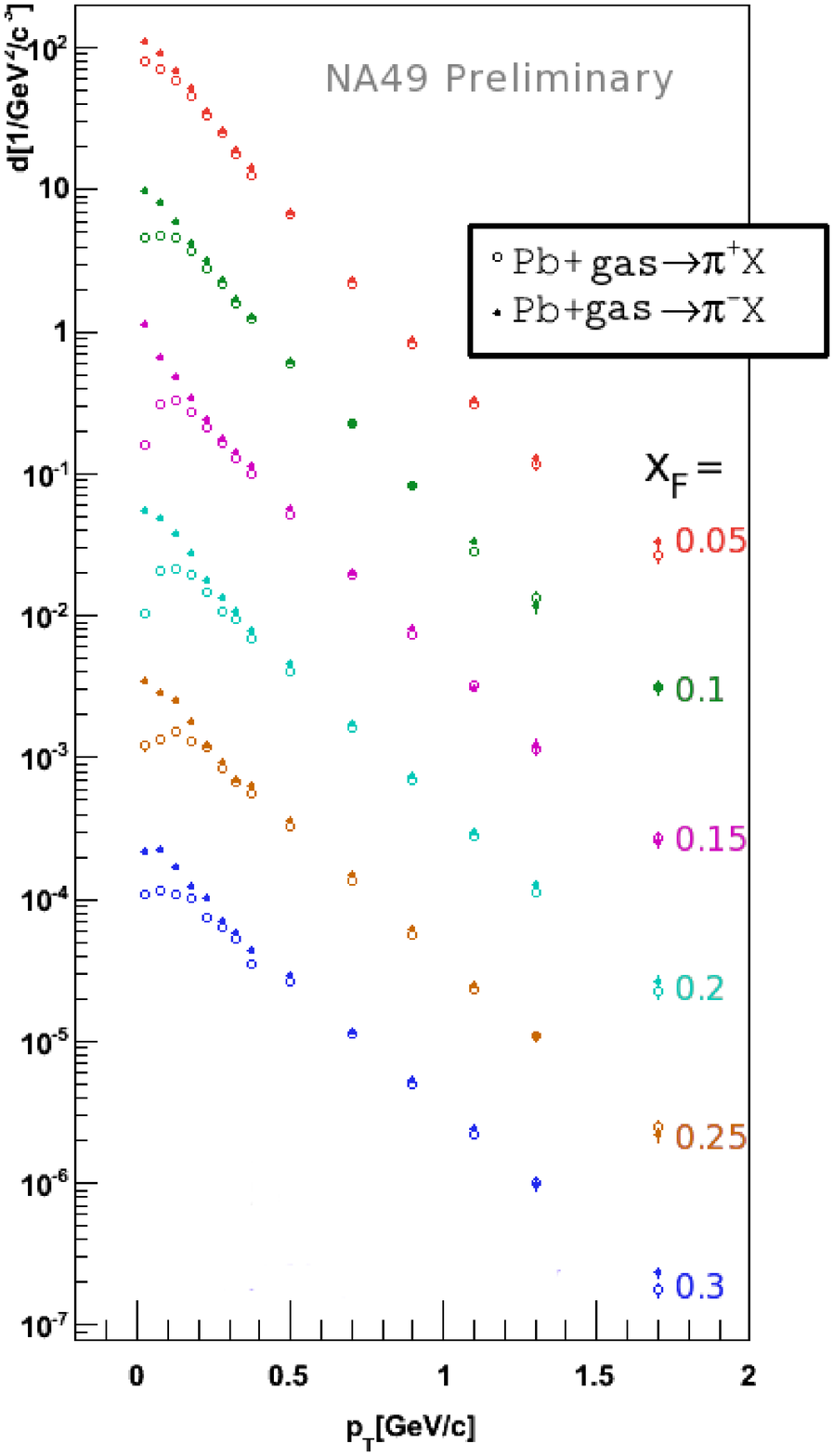}}
 \begin{picture}(10,10)
 \put(79,52){{\bf\Large{(a)}}}
 \put(222,52){{\bf\Large{(b)}}}
 \end{picture}
 \caption{Invariant density $d=E\frac{d^{3}n}{dp^3}$ of positive and 
negative pions produced in (a) peripheral Pb+Pb and (b) Pb+gas 
collisions, drawn as a function of pion transverse 
momentum at fixed values of $x_F$ (listed from the top to the bottom 
distribution). Note: for clarity, the subsequent distributions 
are multiplied by (a)
 $1$, $0.1$, $0.01$, $10^{-3}$, $10^{-4}$, $2\cdot 10^{-6}$, $10^{-7}$, 
$10^{-8}$, $10^{-9}$,
 and by (b) 
 $1$, $0.1$, $0.01$, $10^{-3}$, $10^{-4}$, $10^{-5}$.
 The data come from (a)~\cite{pos,hab} and (b)~\cite{iwonameson}.}  
 \label{ptspectra}
\end{figure}

\section{The data}
 All the studies summarized in this paper were inspired by experimental 
observations made in the SPS energy regime (beam energy of 158 
GeV/nu\-cle\-on, $\sqrt{s_{NN}}=17.3$~GeV). These were coming from two 
data 
sets, both collected by the NA49 experiment~\cite{nim}. The first was a 
sample of peripheral Pb+Pb reactions, defined by a cut of 
150-300 charged particles measured by the NA49 detector~\cite{pos}. This 
corresponded to an average number of 54$\pm$11 participating nucleons 
and a mean impact parameter of 10.9$\pm$0.5 fm~\cite{hab}.
 The second data set was a sample of ``Pb+gas'' events, i.e. collisions 
of the Pb beam projectiles with nuclei of gas surrounding the NA49 
target~\cite{iwonameson}. These collisions (mostly Pb+N, Pb+O) were 
experimentally isolated by means of proper interaction vertex cuts, 
together with the same cut on measured multiplicity as given 
above~\cite{iwonamgr}. It is therefore expected that the corresponding 
mean number of participating nucleons was roughly comparable to that in 
peripheral Pb+Pb reactions. The double differential spectra of 
positively and negatively charged pions produced in both reaction 
samples are shown in Fig.~\ref{ptspectra}. They are drawn as a function 
of transverse momentum $p_T$ at fixed values of the Feynman 
variable\footnote{All the kinematical variables will always be defined 
in the collision c.m.s.} $x_F=\frac{2p_L}{\sqrt{s_{NN}}}$.



\section{The ratios $\frac{\pi^+}{\pi^-}$ in Pb+Pb and Pb+gas events}
\label{three}
\begin{figure}[htb]
\centerline{%
 \includegraphics[width=12.5cm]{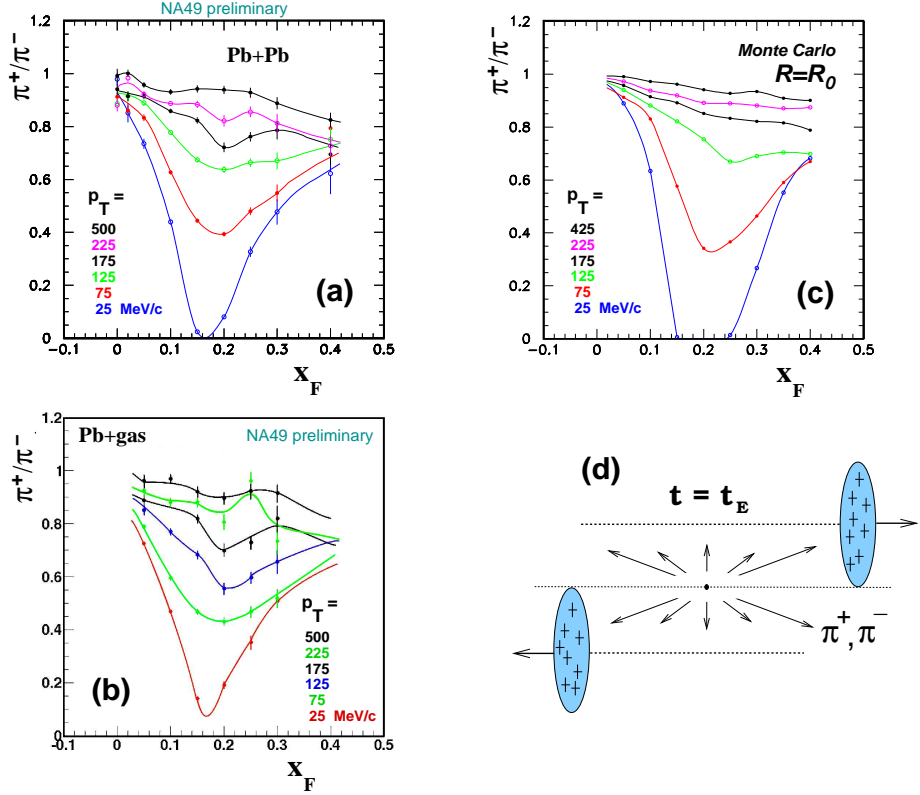}}
 \caption{(a) $\frac{\pi^+}{\pi^-}$ ratio in peripheral Pb+Pb reactions, 
(b) the same ratio in Pb+gas collisions, (c) model simulation as 
described in the text, (d) schematic illustration of the model. The 
data in panel (b) come from~\cite{iwonameson}. 
Other panels come from~\cite{sqm}.}
\label{piratios}
\end{figure}

 Figs~\ref{piratios}(a) and ~\ref{piratios}(b) show the ratios of 
positively 
over 
negatively charged pions produced in peripheral Pb+Pb and Pb+gas 
collisions, respectively. The similarity of the evolution of the ratios 
with $x_F$ and 
$p_T$ in the two collision types is evident. In both cases, the 
$\frac{\pi^+}{\pi^-}$ ratio displays a steep, rapidly varying structure, 
with a deep minimum at low transverse momenta where it comes below 0.2. 
Account taken of the comparable number of protons and neutrons 
participating in both reactions (40\% over 60\% for the Pb nucleus), 
such a low value of this ratio breaks isospin symmetry. This
demonstrates 
that 
the strong interaction cannot be the sole responsible for this effect. 
The position of the minimum ($x_F\approx 0.15=\frac{m_{\pi}}{m_N}$) 
corresponds to pions moving longitudinally with spectator velocity. 
Indeed, the effect is caused by the {\em electromagnetic interaction 
between charged pions and the spectator system}. The repulsion of 
positive pions results in the depopulation of the region $x_F\approx 
0.15$, low $p_T$, while the attraction of negative pions causes their 
accumulation in the same region of phase space (see also 
Fig.~\ref{ptspectra}).

In order to quantify this hypothesis, we constructed an intentionally 
simplified model of the peripheral Pb+Pb collision, taking the 
spectator-induced electromagnetic interaction into 
account~\cite{pos,twospec}. This is illustrated in Fig.~\ref{piratios}(d). 
The two spectator systems are modeled as two Lorentz-contracted, 
uniformly charged spheres. Charged pions are emitted from a single point 
in space (the interaction point). The time of pion emission $t_E$ is a 
free parameter of the model, defining the initial conditions for the 
electromagnetic interaction.
 Initial spectra of pions are assumed to be 
equal to these in the mixture of nucleon-nucleon ($p$+$p$, $n$+$p$, 
$p$+$n$, 
$n$+$n$) 
collisions. They are constructed on the basis of NA49 $p$+$p$ 
data~\cite{pp}.  
Proper account is 
taken 
of isospin 
symmetry
and
of 
the proton/neutron ratio in the Pb nucleus (40\%/60\%);
details of this procedure can be found in~\cite{ondrej}.
 Charged pions are 
then 
numerically propagated in 
the electromagnetic field of the spectators. Relativistic effects such 
as retardation are taken into account.

As it can be seen in Fig.~\ref{piratios}(c), the model gives a 
fair 
description of the main features of the peripheral Pb+Pb data,  
Fig.~\ref{piratios}(a). 
It is to be noted that the simulation presented here assumes a 
mixed set of initial conditions with 50\% of events generated with 
$t_E=0.5$~fm/c and the remaining 50\% generated with $t_E=1$~fm/c. This 
 means
 that the pion formation site is placed ``behind'' the nearest 
spectator system as discussed in~\cite{hab}.
 The only region where a 
more significant disagreement between data and model can be 
seen 
is $x_F\approx 0.2$, low $p_T$. This has been identified as resulting 
from the
{\em fragmentation (break-up) of the spectator system}, and 
yielding in fact
 independent
 information on the space-time evolution of this 
process~\cite{zako}. 

\begin{figure}[htb]
\centerline{%
 \includegraphics[width=12.5cm]{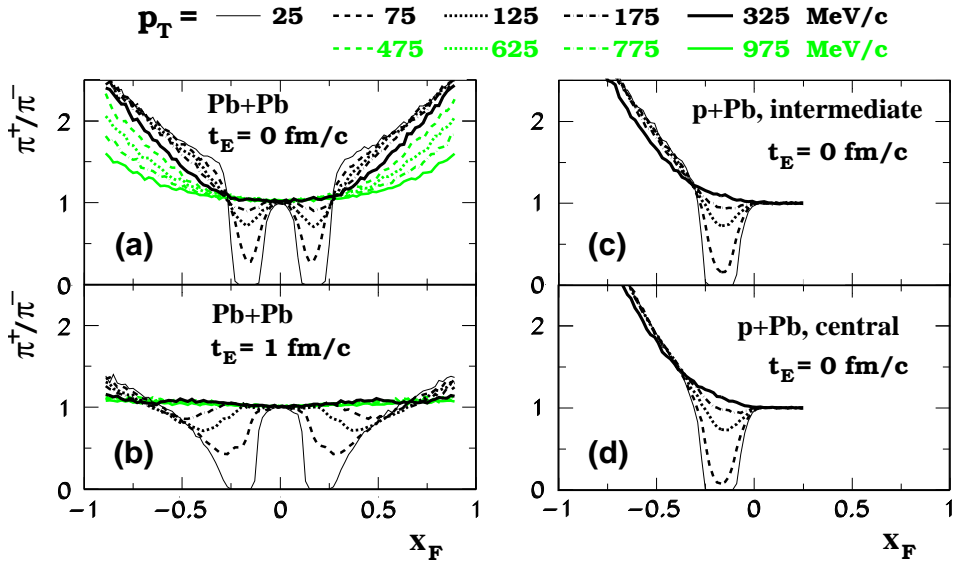}}
\caption{$\frac{\pi^+}{\pi^-}$ ratio in peripheral Pb+Pb reactions 
(a,b), intermediate p+Pb collisions~(c) and central p+Pb collisions (d). 
Note that the simulation of Pb+Pb reactions extends further in $p_T$ 
relative to our 
 earlier work~\cite{twospec}.}
\label{fullxf}
\end{figure}

\section{$\frac{\pi^+}{\pi^-}$ ratios over full phase space}
 Account taken of the success of our model in describing the basic 
features of the experimental data, we decide that it may be applied 
with some confidence to investigate the electromagnetic distortion of 
$\frac{\pi^+}{\pi^-}$ ratios 
 in the full range of $x_F$.
 The results of our simulations\footnote{Note: for
simplicity,
 this
 version of our simulation neglects
 the differences between initial $\pi^+$ and $\pi^-$ 
 spectra,
 resulting from the excess of neutrons in the Pb 
nucleus (these differences were taken into account in the previous 
Section). This 
implies that 
any deviation of the $\frac{\pi^+}{\pi^-}$ ratio from unity will solely 
be 
due to 
electromagnetic effects.}
 are shown in Fig.~\ref{fullxf}. As apparent in 
Fig.~\ref{fullxf}(a),
 the two-dimensional electromagnetic distortion seen in the data in 
Section~\ref{three} is in fact part of a larger structure, consisting of 
two symmetric ``holes'' in the $\frac{\pi^+}{\pi^-}$ ratio, accompanied 
by a 
possible rise at higher absolute $x_F$. It is interesting to note that 
the region of sensitivity to the electromagnetic effect extends, at high 
$x_F$, to relatively large transverse momenta, up to and above 
$p_T=1$~GeV/c.

Comparison with  Fig.~\ref{fullxf}(b) demonstrates the 
sensitivity of the 
spectator-induced electromagnetic distortion to the pion emission time 
$t_E$. This shows the important ability of the electromagnetic effect to 
constrain 
possible scenarios and to provide new information on the space-time 
evolution of the process of pion production, 
see also~\cite{pos,twospec,sqm,zako}.

Finally, the electromagnetic effect in 
intermediate 
and 
central 
p+Pb 
reactions is shown in Figs~\ref{fullxf}(c) and \ref{fullxf}(d). With the 
trivial 
exception of 
the ``one-hole'' structure imposed by the presence of a single 
spectator nucleus, the similarity to peripheral Pb+Pb collisions, 
Fig.~\ref{fullxf}(a), is evident. Little dependence on pure geometry 
(centrality) of 
the p+Pb collision is apparent in the Figure. This gives a relatively 
well defined situation for using the electromagnetic interaction as 
source of information on pion production and on its interplay with 
nuclear fragmentation, also in p+A collisions. 


\section{Conclusions \& outlook}
The spectator-induced electromagnetic effect 
 discussed in this paper
 influences a number of observables in 
various 
types of 
nuclear collisions (p+A, Pb+gas, Pb+Pb, etc). The well-defined nature of 
the electromagnetic interaction makes it a convenient tool to provide 
independent information on the space-time evolution of the reaction. 
However, only a few phenomena have been addressed 
 here.
Other examples can be quo\-ted. Among these, 
the electromagnetic effect induced on kaon spectra~\cite{sqm}, on 
azimuthal anisotropies (directed flow), its 
possible presence 
in ``ultra-peripheral'' Pb+Pb 
collisions~\cite{szczurekchris},
and many others are presently under active investigation.\\

This work was supported by the Polish National Science Centre 
(on the basis of decision no. DEC-2011/03/B/ST2/02634).



\begin{thebibliography}{99}

 \bibitem{pos}
A.~Rybicki, PoS(EPS-HEP 2009) 031.
 %
 \bibitem{hab}
A.~Rybicki, habilitation thesis, Report no.~2040/PH, 
H.~Niewodnicza\'nski Institute of 
Nuclear Physics, Polish 
Academy of Sciences, Krak\'ow, 2010.
{\it http://www.ifj.edu.pl/publ/reports/2010/}
 %
 \bibitem{iwonameson}
I. Sputowska and A. Rybicki,
to appear in Eur. Phys. J. Web. of Conf.
 %
\bibitem{nim}
S.~Afanasev {\it et al.}, NA49 Collab.,
{\it Nucl.\ Instrum.\ Meth.\ {\bf A430}}, 210 (1999).
%
 \bibitem{iwonamgr}
I. Sputowska, M.~Sc.~thesis,
AGH Univ. of Science and Technology, Krak\'ow, 2010.
 See also: 
 A. L\'aszl\'o, Ph. D. thesis, KFKI Research Institute for Particle and 
Nuclear Physics, Budapest, 2007.
 %
 %
 \bibitem{twospec}
  A.~Rybicki, A.~Szczurek, {\it Phys. Rev.} {\bf C75}, 054903 
(2007).
 %
  \bibitem{sqm}
  A.~Rybicki, A.~Szczurek, E.~Kozik, 
{\it Acta Phys. Polon. Supp} {\bf 5}, 369 (2012).
 %
  \bibitem{pp}  
  C.~Alt {\it et al.}, NA49 Collab.,  
  {\it Eur.\ Phys.\ J.} {\bf C45}, 343 (2006).
 %
  \bibitem{ondrej}
  O.~Chvala (NA49 Collab.),
  {\it Eur.\ Phys.\ J.} {\bf C33}, S615 (2004).
 %
 \bibitem{zako}
A.~Rybicki, {\it Acta Phys. Polon.} {\bf B42}, 867 (2011).
 %
 \bibitem{szczurekchris}
 M. K{\l}usek-Gawenda and A. Szczurek,
 {\it Phys. Rev.} {\bf C82}, 014904 (2010).
\end{thebibliography}
\end{document}